\documentclass[aps,prl,twocolumn,superscriptaddress]{revtex4}%
\usepackage{subfigure}
\usepackage{amsmath}
\usepackage{graphicx}
\usepackage{rotating}
\usepackage{amsfonts}
\usepackage{amssymb}%
\providecommand{\U}[1]{\protect\rule{.1in}{.1in}}
\begin{document}
\title{Unconventional sign-reversing superconductivity in LaFeAsO$_{1-x}$F$_{x}$}
\author{I.I. Mazin}
\affiliation{Code 6393, Naval Research Laboratory, Washington, D.C. 20375}
\author{D.J. Singh}
\affiliation{Materials Science and Technology Division, Oak Ridge National Laboratory, Oak
Ridge, TN 37831-6114}
\author{M.D. Johannes}
\affiliation{Code 6393, Naval Research Laboratory, Washington, D.C. 20375}
\author{M.H. Du}
\affiliation{Materials Science and Technology Division, Oak Ridge National Laboratory, Oak
Ridge, TN 37831-6114}
\date{Printed \today }

\begin{abstract}
We argue that the newly discovered superconductivity in a nearly magnetic,
Fe-based layered compound is unconventional and mediated by antiferromagnetic
spin fluctuations, though different from the usual superexchange and specific
to this compound. This resulting state is an example of extended $s$-wave
pairing with a sign reversal of the order parameter between different Fermi
surface sheets. The main role of doping in this scenario is to lower the
density of states and suppress the pair-breaking \textit{ferromagnetic} fluctuations.

\end{abstract}
\maketitle

The discovery of superconductivity with $T_{c}\gtrsim$ 26 K \cite{JACS} in a
compound  that contains doped Fe$^{2+}$ square lattice sheets raises immediate
questions about the nature of the superconducting state and the  pairing
mechanism. A number of  highly unusual properties suggest, even at this early
stage, that  the superconductivity is not conventional. We argue that not only
is it unconventional,  but that doped LaFeAsO represents the first example of
multigap superconductivity with a discontinuous  change of the order parameter
(OP) phase between bands, a state discussed previously ($e.g.,$
Refs.\cite{MazinGolubov,GorkovAgterberg}), but not yet observed in nature.

We suggest that superconductivity here is mediated by spin fluctuations (SF),
as many believe is the case in cuprates, heavy fermion materials, or
ruthenates. SF can only induce a triplet superconducting OP, or a singlet one
that changes sign over the Fermi surface (FS). The latter condition often, but
not always, dictates strong angular anisotropy of the OP (cf. d-wave). In 
our
scenario it is satisfied despite full angular isotropy, since the sign
changes between the two sets of FSs. Our model is also principally
different from the convetional s-wave superconductivity discovered in
MgB$_{2}$: there the pairing interaction is attractive, in our case it is
repulsive (but pairing thanks to the sign reversal). Finally, similar to d- or
p-wave superconductivity our OP has a nearest-neighbor structure in the real
space, thus reducing the Coulomb repulsion within the pair.

We begin by arguing against conventional superconductivity. The pure compound,
LaFeAsO, is on the verge of a magnetic instability: it has a very high
magnetic susceptibility\cite{JACS} and is strongly renormalized 
compared to
density functional (DFT) calculations \cite{singh}. This renormalization is
higher than in any known conventional superconductor, including MgCNi$_{3}$,
where superconductivity is believed to be strongly depressed by spin
fluctuations.  The susceptibility in the pure compound is large and upon
doping with F grows even larger and becomes Curie-Weiss like. This suggests
nearness to a critical point in the pure compound and non-trivial competition
between different  spin fluctuations (SF). Very strong electron phonon
interactions would be required to overcome the destructive effects of such SF.
We have calculated \textit{ab initio} the electron-phonon spectral function,
$\alpha^{2}F(\omega)$, and coupling, $\lambda$, for the stoichiometric
compound \cite{espresso}. Some moderate coupling exists, mostly to As modes,
but the total $\lambda$ appears to be $\sim0.2$, with $\omega_{log} \sim250$K ($\omega_{log}$ is the 
logarithmically averaged boson frequency of Eliashberg theory), which  can in no way explain $T_{c} \gtrsim26$ K.

\begin{figure}[ptb]
\includegraphics[width = 0.95\linewidth]{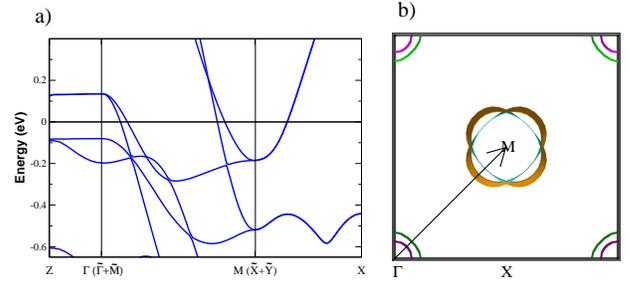}
\caption{{color online}(a)
Calculated band structure at $x=0.1$ near the Fermi Level. (b) Calculated
Fermi surface at 10\% F doping. Note that the only 3D
parts are the far ends of the electron cylinders around M. The fully three-dimensional surface 
present in the undoped compound is suppressed beneath E$_F$ by increased electron count.}
\label{FS}
\end{figure}

The calculated DFT Fermi surfaces \cite{singh} for undoped LaFeAsO consist of
two small electron cylinders around the tetragonal $M$ point, and two hole
cylinders, plus a heavy 3D hole pocket around $\Gamma$. To study doping
effects, we performed full-potential calculations using the WIEN package in
the virtual crystal approximation and the PBE GGA functional. The lattice
parameters we took from experiment \cite{JACS} and optimized the internal
positions \cite{positions}. In Fig. \ref{FS} we show the bands near the Fermi
level for $x=0.1$ and the corresponding Fermi surface. As expected, the 3D
pocket fills with electron doping (at $x=0.04$-0.05) and the 
fermiology
radically simplifies, leaving a highly 2D electronic structure with two heavy
hole cylinders and two lighter (and larger) electron cylinders.

\begin{figure}[ptb]
\includegraphics[width = 0.99\linewidth]{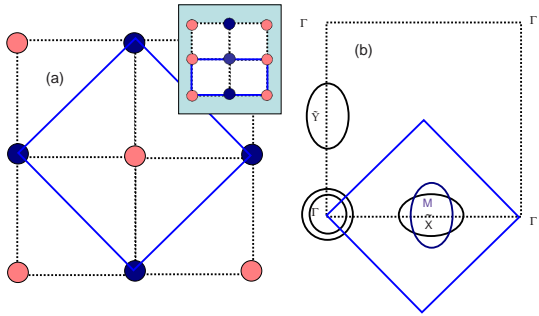} \caption{{color online}
Fermi surface formation upon backfolding of the large BZ
corresponding to a simple Fe square lattice. (a) Real space: the four small
unit cells of the Fe sublattice only (dashed) with the larger solid diamond of
actual two-Fe unit cell. Dark and light circles indicate superexchange
(checkerboard) ordering. The inset shows the spin density wave corresponding
to \~X point SF. (b) Reciprocal space: the black square is the unfolded
BZ, the blue diamond is the downfolded zone, the blue ellipse is
the electron pocket from the \~Y point downfolded onto the \~X point (which is
M in the small BZ).}
\label{bz}
\end{figure}This fermiology imposes strong constraints on the
superconductivity. In particular, with the exception of phonons, it is hard to
identify pairing interactions with a strong $k_{z}$ dependence. Thus, states
with strong variations of the OP along $k_{z}$ are unlikely. An angular
variation of the OP in the $xy$ plane is possible, but would require an
unrealistically  strong $\mathbf{q}$-dependence of the pairing interaction on
the scale of the small Fermi surface size, and would also be extremely
sensitive to impurities.

On the other hand, the small Fermi surfaces are readily compatible with a
pairing state with weak variations of the OP within the sheets, but a $\pi$
phase shift between electron and hole cylinders. Here we show that a SF
pairing interaction favoring exactly such a state is present in this material
and we discuss the expected consequent physical properties.

The SF spectrum is unusually rich for this compound and comes from three
separate sources. First, the system is relatively close to a Stoner
ferromagnetic instability. Second, there is a nearest-neighbor
antiferromagnetic (AFM) superexchange\cite{added}(d). Third, there are
nesting-related AFM spin-density-wave type SF near wave vectors connecting the
electron and hole pockets. The latter appear to be the strongest ones.
The corresponding interaction connects the well-separated FS pockets located
around $\Gamma$, and around $M$. Though repulsive in the singlet channel,
these would nevertheless be strongly pairing provided that the OPs on the two
sets of the FSs have opposite signs. \textit{The main message of our paper
is that this \textquotedblleft$s_{\pm}$\textquotedblright\ superconducting
state is both consistent with experimental observations and most favored by
SF in this system}.

As opposed to the undoped material\cite{singh}, we do not find any FM solution
for the doped compound, even with the more magnetic GGA functional. This
suggests that \textit{the main function of doping is to move the system away
from a ferromagnetic instability} (see, however, Ref. \cite{added}). That the
system becomes less magnetic is due to the fact that upon doping the heavy 3D
hole pocket near $\Gamma$ rapidly fills and the total DOS drops (by a factor
of two). At a doping level $x=0.1$, we obtain in the GGA an unrenormalized
Pauli susceptibility $\operatorname{Re}\chi_{0}(\mathbf{q}=0,\omega
=0)\approx4\times10^{-5}$ emu/mole ($N(0)=0.64$ states/eV/spin/Fe). Using 1.1
eV for the Stoner $I$ on Fe, we obtain a renormalized $\chi(0)$ of
0.14$\times10^{-3}$ emu/mole, a large renormalization, but much smaller than
what is needed to explain the experimental value\cite{JACS}. Note that in the
undoped system the calculated susceptibility is larger, and the experimental
one smaller, than in the doped one. This suggests that besides FM SF there are
other, more important spin excitations in the system.

The first candidate for these is a superexchange corresponding to the standard simple 
nearest-neighbor checkerboard antiferromagnetism. Importantly, there is also substantial direct 
Fe-Fe overlap\cite{singh}, which leads to an additional AFM exchange of comparable strength and 
with the same checkerboard geometry.

Further discussion requires an understanding of the fermiology in clearer terms. The band 
structure may at first appear intractably complex, but it is in fact relatively simple, involving 
only three Fe orbitals near the Fermi level. The hole pockets around $\Gamma$ originate from Fe 
$d_{xz}$ and $d_{yz}$ bands that are degenerate at $\Gamma,$ and form two nearly perfect 
concentric cylinders. The electron surfaces are better understood if we recall that the 
underlying Fe layer forms a square lattice with the period $\tilde {a}=a/\sqrt{2}$ (Fig. 
\ref{bz}). The corresponding 2D Brillouin zone (BZ) is twice larger and rotated by 45$^{\circ}.$ 
If we could \textquotedblleft unfold\textquotedblright\ the Fermi surface of Fig. \ref{FS}, we 
would find the same two hole pockets at $\tilde{\Gamma}$ and $one$ electron pocket around the 
point \~{X} in the \textit{large} BZ. The latter is formed by the $d_{yz}$ band (or $d_{xz}$ near 
\~{Y}) that starts 180 meV below the Fermi energy, disperses up along \~{X} \~{M} and is 
practically flat along \~{X} $\tilde{\Gamma}$. This band is hybridized with the $d_{xy}$ band (or 
$d_{x^{2}-y^{2}}$ in the two-Fe cell), which starts from \~{X} at an energy of -520 meV, and is 
instead dispersive along \~{X} $\tilde{\Gamma}$ . Upon hybridization, these two bands yield an 
oval cylindrical electron pocket, elongated along $\tilde{\Gamma}$\~{X} \cite{notelee}.

Electronic transitions from the hole pockets to the electron pocket should
lead to a broad (because the electron pocket is oval rather than circular)
peak in the \textit{noninteracting }susceptibility $\chi_{0}(q,\omega
\rightarrow0)$, at $q=(\pi/\tilde{a},0)$, while the superexchange interaction
$J(q)$ on the square lattice should be peaked at $q=(\pi/\tilde a,\pi/\tilde a)$ ($\Gamma$
in the downfolded BZ). The renormalized susceptibility, $\chi(q)=\chi
_{0}(q)/[1-J(q)\chi_{0}(q)]$ then has a rich structure with maxima at
$\tilde{\Gamma},$ \~{X} and \~{M}. For the true unit cell with two Fe, both
$\tilde{\Gamma}$ and \~{M} fold down into the $\Gamma$ point, while \~{X}
folds down into the M point. The corresponding folding of the Fermi surfaces
makes the electron pockets overlap, forming two intersecting surfaces (Fig.
\ref{FS}). It is important to appreciate from this \textit{gedanken unfolding
} that already on the level of the noninteracting susceptibility there is a
tendency for antiferromagnetic correlations with a wave vector different from
the superexchange one.

\begin{figure}[ptb]
\includegraphics[width = 0.95\linewidth]{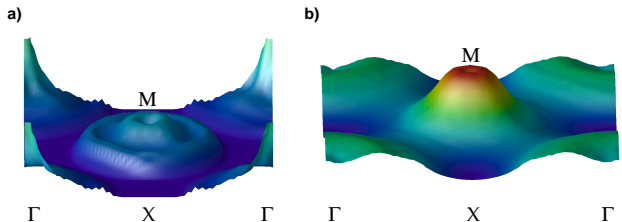}\caption{{color online} The
imaginary (a) and the real (b) parts of the non-interacting susceptibility
$\chi_{0}(q,\omega\rightarrow0)$, in arbitrary units. Within common
approximations Im$\chi$=Im$[\frac{\chi_{0}}{(1-J(\mathbf{q})\chi_{0}(q,w)}]$
is measurable by neutron scattering and Re$\chi$ controls the pairing
interaction, in the singlet channel proportional to $1/[1-J(\mathbf{q}
)\chi_{0}(q,0)]$ (see Ref. \cite{MU} for a review.) Note that the RPA
enhancement will strengthen both peaks.}
\label{chi}
\end{figure}

In Fig. \ref{chi} we present the calculated \cite{chi}  $\chi_{0}(q,\omega)$ =
$\frac{f(\epsilon_{k})-f(\epsilon_{k+q})}{\epsilon_{k}-\epsilon_{k+q}%
-\omega-i\delta}$ at $\omega\rightarrow0$ and $q_{z}=\pi/c$ ($\chi$ is
practically independent of $q_{z}$), The peak at M, derived from interband
transitions, is very broad, as expected, with some structure around the M
point, due to the particular orientation of the two oval pockets at M and the
size difference between the hole and electron cylinders for finite doping.  With minor modification, this structure is present also in the undoped 
compound.

We have also performed magnetic calculations (discussed in a separate
publication) in a supercell  corresponding to the
superexchange, $\mathbf{q}=$ \~{M} (=$\Gamma$ in the downfolded Brillouin
zone) and nesting-induced $\mathbf{q}=$ \~{X} (=M) spin density waves. We
find, indeed, that the tendency to ferromagnetic ordering is suppressed in the
doped compound, while the tendency to nesting-based antiferromagnetism is
present, even leading to an actual instability at the mean field level.  It is worth noting that while this instability does not appear in actual 
superconducting materials, it has been now found experimentally in the low doping regime \cite{added}.

The strong AFM SF around M favor our proposed $s_{\pm}$ state. Cases where
SF-induced interactions connect two pockets of the Fermi surfaces, including
SF originating from electronic transitions between the very same pockets have
been considered in the past. \cite{odd} However, they involved FS pockets
related by symmetry, which strongly restricts the phase relations between
them. In our case the two sets of pockets are not symmetry
related, and nothing prevents them from assuming arbitrary phases.

For the singlet case, the coupling matrix between the hole ($h)$ and the
electron pockets ($e)$ is negative, $\lambda_{eh}<0.$ The diagonal components
emerge from competition between the attractive phonon-mediated and repulsive
SF-mediated interactions and are, presumably, weak. If $\lambda_{hh}^{ph}$ is
the average of the phonon-mediated interaction over the wave vectors
$q<0<2k_{F}^{h}$, and $\lambda_{hh}^{sf}$ the same for the SF, then
$\lambda_{hh}=\lambda_{hh}^{ph}-\lambda_{hh}^{sf}.$ Our calculated
electron-phonon coupling comes mostly from small wave vectors, that is,
$\lambda_{hh}^{ph}+\lambda_{ee}^{ph}\approx0.2$, and $\lambda_{eh}^{ph}%
\approx0$. Phonons, therefore, though weak, promote the $s_{\pm}$ state. On
the other hand, $\lambda_{hh}^{sf}$ and $\lambda_{ee}^{sf}$, come from FM
(small-q) SF and are pair breaking. Finally, the superexchange interaction
does not affect $\lambda_{hh},$ but the corresponding wave vector
$\mathbf{q}=\tilde{\Gamma}$\~{M} connects the $e$ pockets near \~{X} and \~{Y}
and are also pair breaking\cite{added}(c).

The $T_{c}$, as usual for a two-gap superconductor\cite{MazinAntropovReview},
is defined by the maximal eigenvalue of the $\lambda$ matrix, $\lambda
_{eff}=[\lambda_{hh}+\lambda_{ee}+\sqrt{4\lambda_{eh}\lambda_{he}%
+(\lambda_{ee}-\lambda_{hh})^{2}}]/2,$ and the OPs $\Delta_{e,h}$ are defined
by the corresponding eigenvector. In our case, the signs of $\Delta_{h}$ and
$\Delta_{e}$ will be opposite, and their absolute values (despite the
different densities of states) will be similar since presumably $(\lambda
_{ee}-\lambda_{hh})^{2}\ll\lambda_{eh}\lambda_{he}.$ This \textquotedblleft%
$s_{\pm}$\textquotedblright\ state is analogous to the states proposed
previously for semimetals\cite{AronovSonin} and bilayer
cuprates\cite{LichtMazin}.

From the point of view of neutron scattering, the structure of the peak in
$\chi$ near M is important, but for SF induced superconductivity it does not
matter at all. A well defined spin-excitation requires a sharp peak, but the
pairing interaction is integrated over all possible $q$ vectors spanning the
two sets of Fermi surfaces so that only the total weight of the peak is
important. Dimensionality is however very important, as 3D coupling supports a
long range magnetic ordering (competing with superconductivity). Indeed, the
experiment shows an abrupt appearence of superconductivity at $x\approx0.03,$
roughly where the 3D pockets around $\Gamma$ disappear.

One might envision a triplet state similar to the described singlet one, fully
gapped, as expected for unitary 2D $p$-wave states, and with different
amplitudes (or signs) on the two cylinders. The similarity,
however, is misleading. In the triplet channel, SF induce attraction, but
given the relatively large width of the AF peak (Fig.\ref{chi}), a large part
of the pairing will be lost as only SF with a wave vector exactly equal to
($\pi/a,\pi/a)$ will be fully pairing, and some others will even be
pair-breaking. Therefore, in this scenario where the antiferromagnetic SF
around the $M$ point provide the primary pairing interaction, we expect the
lowest energy superconducting state to be $s_{\pm}$. The structure of the OP
in real space can be evaluated using the lowest Fourier component,
compatible with the proposed  $s_{\pm}$ state, namely $\cos k_{x}+\cos k_{y}$
(equivalently, $\cos\widetilde{k_{x}}\cos\widetilde{k_{y}}).$ This 
corresponds
to pairing of electrons that reside on the nearest neighbors, just as in 
the
d-wave case, so that the onsite Coulomb repulsion is not particularly
destructive for this state.

Finally, we discuss the experimental ramifications of the $s_{\pm}$ state. In 
many aspects they are similar to that of the $s_{\pm}$ state proposed in Ref. 
\onlinecite{LichtMazin} and discussed in some detail in 
Refs.\cite{LichtMazin,MazGZ,MazY}. The thermodynamic and tunneling 
characteristics are the same as for a conventional two-gap 
superconductor\cite{MazinAntropovReview}. The two-gap character, however, may 
be difficult to resolve, given the dominance of the interband interactions 
which will render the two gap magnitudes similar. Nonmagnetic small-$q$ 
intraband ($e-e$ or $h-h)$ scattering, as well as the interband spin-flip 
pairing will not be pair breaking, but paramagnetic interband scattering will, 
resulting in finite DOS below the gap\cite{MazinGolubov}, consistent with 
specific heat measurements \cite{heat}. The most interesting feature of the 
$s_{\pm}$ state \cite{MazY} is that the coherence factors for exciting 
Bogolyubov quasiparticles on FS sheets with opposite signs of the OP are 
reversed compared to conventional coherence factors. We outline a few 
important consequences of the relatively straigtforward application of this 
concept to experimental probes. First, one expects a qualitative difference 
between experiments that probe vertical transitions (\textbf{q}=0) and those 
that probe transitions with \textbf{q} close to $\pi/a,\pi/a.$ For instance, 
the spin susceptibility at \textbf{q}=0 will behave conventionally, 
\textit{i.e.} exponentially decay below $T_{c}$ without any coherence peak, 
while the susceptibility for \textbf{q}$\approx\pi/a,\pi/a$ will have a 
coherence peak that should be detectable by neutron scattering as an 
enhancement below $T_{c}$ \cite{MazY}. AFM SF near $\pi/a,\pi/a$ dominate in 
the doped material, and so the usual coherence peak in the NMR relaxation 
rate, which averages equally over all wave vectors is expected to disappear or 
be strongly reduced. It was shown in Ref. \cite{MazGZ} that Josephson currents 
from FSs with different signs of the OP interfere destructively, and the net 
phase corresponds to the sign of the FS with the higher normal conductance. In 
the constant relaxation time approximation, both in-plane and out of plane 
conductivities are dominated by the electron pockets. This is unfortunate, 
since there would otherwise be a $\pi$ phase shift between the $ab$ and $c$ 
tunneling and corner-junction experiments could be used, as in high-T$_{c}$ 
cuprates.

To summarize, we argue that the fermiology found in doped LaAsFeO gives rise
to strong, but broad antiferromagnetic spin fluctuations near the M point in
the Brillouin zone, while the tendency to magnetism existing at zero doping is
suppressed. These fluctuations, while too broad to induce a magnetic
instability, are instrumental in creating a superconducting state with OPs of
opposite signs on the electron and hole pockets. 

We would like to acknowledge helpful discussions with D. Scalapino. Work at
NRL is supported by ONR. Work at ORNL was supported by DOE, Division of
Materials Science and Engineering.


\begin{thebibliography}{99}
\bibitem {positions}The optimized positions actually depend slightly on
doping. For $x=0.1$, in GGA, they are: $z_{La}=0.148$, $z_{As}=0.638$.
\bibitem {chi}We used a three-dimensional grid of $\approx$ 75,000 $k$ and $q$
points with a temperature smearing of 1 mRy, and the constant matrix element approximation
\end{thebibliography}
\end{document}